\documentclass[aps, prx, twocolumn, superscriptaddress, longbibliography,10pt]{revtex4-2}

\usepackage{graphicx}
\usepackage{amsmath}
\usepackage{amssymb}
\usepackage{amsthm}
\usepackage{bbold}
\usepackage{pifont}
\usepackage{array}
\usepackage{tabularx}
\usepackage{xcolor}
\usepackage{psfrag}
\usepackage{soul}
\usepackage{textcomp}
\usepackage{bm}
\usepackage{bbm}
\usepackage{yfonts}
\usepackage{mathtools}

\usepackage{lipsum}

\usepackage[utf8]{inputenc}
\usepackage[english]{babel}
\usepackage[T1]{fontenc}
\usepackage{dcolumn}

\usepackage{tikz}
\usetikzlibrary{quantikz2}

\newcommand{\algo}{AQUIRE}

\let\oldpercent\%
\renewcommand{\%}{\scalebox{0.85}{\oldpercent}}

\usepackage{hyperref}
\hypersetup{
 pdftitle = {An Error-aware and Adaptive Method for the Estimation of Quantum Observables on Qudit-Based Quantum Computers},
 pdfauthor = {R. Simon, A. Jena, L. Dellantonio},
 breaklinks=true,
 pdfnewwindow=true,
 colorlinks=true,
 linkcolor=red,
 citecolor=blue,
 filecolor=blue,
 urlcolor=blue
}

\begin{document}

\title{An Error-aware and Adaptive Method for the Estimation of Quantum Observables on Qudit-Based Quantum Computers}


\author{Rick P. A. Simon}
\email{r.p.a.simon@exeter.ac.uk}
\affiliation{Department of Physics and Astronomy, University of Exeter, Stocker Road, Exeter EX4 4QL, UK}

\author{Michael Meth}
\email{michael.meth@uibk.ac.at}
\affiliation{Universit\"at Innsbruck, Institut f\"ur Experimentalphysik, Technikerstraße 25/4, Innsbruck, Austria}

\author{Francesco Martini}
\affiliation{Department of Physics and Astronomy, University of Exeter, Stocker Road, Exeter EX4 4QL, UK}

\author{Peter Tirler}
\email{peter.tirler@uibk.ac.at}
\affiliation{Universit\"at Innsbruck, Institut f\"ur Experimentalphysik, Technikerstraße 25/4, Innsbruck, Austria}

\author{Andrew Jena}
\email{ajjena@uwaterloo.ca}
\affiliation{softwareQ Inc., Waterloo, Ontario, N2L 0C7, Canada}

\author{Martin Ringbauer}
\email{martin.ringbauer@uibk.ac.at}
\affiliation{Universit\"at Innsbruck, Institut f\"ur Experimentalphysik, Technikerstraße 25/4, Innsbruck, Austria}

\author{Luca Dellantonio}
\email{l.dellantonio@exeter.ac.uk}
\affiliation{Department of Physics and Astronomy, University of Exeter, Stocker Road, Exeter EX4 4QL, UK}

\begin{abstract}
The accurate estimation of observables is a crucial task in quantum computing. Recent advances have highlighted the need for (a) specialized protocols for qudit-based devices, that include (b) error-aware strategies.
Here, we present {\algo}, the first protocol that can (a) accurately estimate both the mean and the error of an observable on qudit-based quantum computers. {\algo} achieves this by constructing a Bayesian model to accommodate generalized Pauli operators. It is designed to continuously monitor the estimated average and the associated error of the observable, adjusting the subsequent measurements in real-time. Additionally, {\algo} is (b) device- and experiment-specific error-aware, and accounts for hardware imperfections and experimental noise during the estimation process.  We demonstrate {\algo}'s advantage via numerical simulations and showcase its ability to quantify the noise affecting the estimation by implementing it on a trapped-ion qudit quantum processor. By exploiting general commutation relations and overlap grouping measurements, our protocol is state-of-the-art when restricted to qubit-based quantum computers and extends this advantage to the qudit case.

\end{abstract}

\maketitle
\section{Introduction} \label{sec:Introduction}
Within the NISQ era and progressing towards early fault-tolerant devices \cite{Aharonov1999,Kitaev2003,Roffe2019,Preskill2021}, quantum computing \cite{Shor1997,PAUL2007,Nielsen2012,Preskill2018,Acin2018} holds the potential to enhance a broad spectrum of applications, including finance \cite{Chakrabarti2021,Braine2021,Huber2024,Martini2025}, logistics \cite{Rieffel2014,Rosenberg2016,Feld2019,Phillipson2024}, chemistry \cite{Cao2018,Cao2019,Motta2021,Mishmash2023,Richerme2023,Maurizio2025}, material science \cite{Kitai2020,Ma2020,Camino2023,Alexeev2024} and high-energy physics \cite{Banuls2020,Haase2021,Paulson2021,Meth2023}. 
The development of the first qudit-based quantum computers (QC) \cite{Ringbauer2022,Chi2022,Low2025} hold the potential to accelerate the fields towards achieving (useful) quantum advantage \cite{Kim2023,Arute2019,King2024,Zhong2020}.
These devices offer the opportunity to work with a larger coding space and a significantly reduced number of qudits.
This is particularly relevant for quantum simulation, one of the most promising applications of QC \cite{Lloyd1996,Georgescu2014,Daley2022}.
However, as of today, there is no practical protocol for the measurement of observables on qudit-based devices. 

Furthermore, despite the progress towards fault-tolerant QC \cite{Bluvstein2024,Ott2024,Qiao2025,Kim2023,Meth2023,Delaney2024,Meng2024}, hardware noise is still a phenomenal obstacle to achieve useful quantum advantage \cite{Preskill2021}. 
While several error mitigation techniques exist \cite{Neeley2010,Bialczak2010,ARahman2022,Atas2022}, none (to our knowledge) is integrated within the measurement protocol to yield an accurate error estimate that is inclusive of both the statistical and hardware noise contributions.
This poses a stringent limitation to all quantum algorithms that must be able to assess whether two quantum states have, e.g., the same energy. 
For instance, variational algorithms \cite{McClean2016,Cerezo2021,Ferguson2021} would greatly benefit from detailed knowledge of the error (inclusive of both statistical and hardware contributions) affecting the state, as this information enables the classical optimizer to avoid allocating measurements to observables with high noise levels.
In this work, we introduce {\algo} (\textbf{A}daptive \textbf{Qu}antum Measurements w\textbf{i}th \textbf{R}eal-time \textbf{E}rror-awareness), the first qudit, error-aware measurement protocol. 

{\algo} allows the user to measure observables for any combination of different-dimensional qudits, all while monitoring the error, both from statistical and hardware noise. The estimation is performed adaptively and relies on a Bayesian approach \cite{Kruschke2014,Gelman1995} that enhances the estimation accuracy. At the same time, {\algo} is not restricted to bitwise commutation and permits overlap between groups of observables that are measured independently.
As demonstrated both by our numerical simulations and by experimental results on trapped-ion qudit QC, {\algo} outperforms the current best measurement protocol for qubits \cite{Shlosberg2023} and applies this advantage to qudit-based QC.

Our work is structured as follows. First, we provide an overview in Sec.~\ref{Sec:Overview}. The main theoretical results, inclusive of the Bayesian estimation model, are in Sec.~\ref{sec:Theory}. Sec.~\ref{sec:Algorithm} describes the algorithmic pipeline for measuring qudit observables. Numerical and experimental results using chemistry and LGT Hamiltonians are in Secs.~\ref{sec:NumResults} and \ref{sec:ExpResults}, respectively. In the latter, we also present our method to include the hardware noise into the error estimate. We conclude in Sec.~\ref{Sec:Conclusion} with open problems and future directions of research.

\section{Overview of the Main Results} \label{Sec:Overview}
Quantum algorithms are based on multiple subroutines, one of which is ubiquitous: the measurement of quantum states. Whether the algorithm is qubit- \cite{Mosca2008} or qudit-encoded \cite{Chi2022}, noisy or fault-tolerant \cite{Gottesman1998}, variational or Grover-based \cite{Kerzner2024}, it is essential to provide the best possible measurement protocol \cite{Shlosberg2023,Vankirk2024,Yen2023,Elben2023}.
For a user-specified accuracy, {\algo} requires the lowest number of measurements $M$ (Sec.~\ref{sec:NumResults}) and provides the circuits required for the Pauli diagonalization (Sec.~\ref{sec:Algorithm}).
Additionally, as demonstrated by Fig.~\ref{Fig:NumericalResults}(a), it accommodates for both bitwise and general commutation (meeting any experimental requirement) and is adaptive (already measured data guides the allocation strategy). 
As output, {\algo} \emph{directly} yields the estimated error (in addition to the average), and thus does not need to acquire (expensive) additional data to assess the measurement accuracy, nor does it rely on error bounds that can greatly overestimate the error \cite{Shlosberg2023}. 

{\algo} is designed for fault-tolerant QC as well as NISQ devices. Not only does it provide an estimate of the statistical error (Sec.~\ref{sec:Theory}), it also quantifies the hardware noise that has directly affected the experimental outcomes. As we discuss in Sec.~\ref{sec:ExpResults}, this requires a limited cost in terms of additional resources, at most comparable with the budget allocated for the measurements.

Finally, {\algo} is qudit ready, promptly adapting to the characteristics of any QC \cite{Meth2023,Ringbauer2022}. Whether the considered observable is qubit-, qudit-, spin-based, or even mixed, our work provides a pipeline with all features discussed above, which we consider necessary for a fully functional measurement protocol.
Based on the numerical and experimental evidence presented below, {\algo} is the state-of-the-art for digital quantum simulations. It yields more accurate results in scenarios [qubit observables, Fig.~\ref{Fig:NumericalResults}(a)] in which it can be benchmarked against other existing protocols and extends that advantage to qudit-based hardware.

For qudit observables, we test {\algo} with the U(1) Lattice Gauge Theory (LGT) Hamiltonians \cite{Meth2023}, which includes particles (qubits) and gauge fields (qutrits and ququints, depending on truncation \cite{Haase2021,Paulson2021}). 
Noiseless numerical simulations in Fig.~\ref{Fig:NumericalResults}(b) show the convergence of the estimated error (dots) to the true error (squares, obtained from the true covariances, unknown in the experiment).
We witness significant advantages in all cases when employing adaptive allocation and allowing general commutation (in contrast to non-adaptive allocation and bitwise commutation). In particular, we see that for many measurements, {\algo} allocates the shots \emph{as if} the quantum state was known \emph{before} the process (black line).

Additionally, we benchmark {\algo} against AEQuO \cite{Shlosberg2023}, that is (to our knowledge) state-of-the-art for qubit-based QCs. 
As can be seen in Fig.~\ref{Fig:NumericalResults}(a), we find that {\algo} outperforms AEQuO when applied to qubit observables.
We note that this is not only in the estimated errors (dots), but also the true errors (squares).
The improved Bayesian estimation strategy, presented in Sec.~\ref{sec:Theory}, allows for more accurate shot allocation and therefore better errors.

For NISQ and error-mitigated QCs, {\algo} monitors the hardware noise that directly affects the measurement. 
We extended the Bayesian estimation strategy (discussed above and in Sec.~\ref{sec:Theory}) to quantify the deviation of the estimated average from the true, unknown value. {\algo} not only recognizes the extent of the noise affecting the apparatus, but also its impact on the measurement (e.g., a noisy measurement of $\ket{+}$ yields a correct estimate of $\langle Z \rangle = 0$). Furthermore, the cost of error awareness is contained. While being ultimately user-specified, in this work we split the measurement budget equally between the estimations of the considered observable and the hardware noise, respectively. 

The main result of this work is Fig.~\ref{Fig:ExpData}, where we demonstrate the performance of {\algo} on the trapped-ion qudit QC in \cite{Ringbauer2022}.
Figs.~\ref{Fig:ExpData}(a-b) demonstrate that {\algo} obtains meaningful estimation errors for an experiment in a noisy environment. Figs.~\ref{Fig:ExpData}(c-d) showcase how to achieve the highest possible accuracy, limited by the characteristics of the hardware noise that are directly detrimental to the considered observable, unknown quantum state, and measurement protocol. 
As expected, while entangling gates allow to gather more data per measurement shot, their increased error rate limits the best achievable estimation accuracy (if compared to using local gates only). This is shown in Fig.~\ref{Fig:ExpData}(d), suggesting that a good strategy is to allow for non-bitwise commuting terms to be simultaneously probed only in an initial phase of the measurement, while the number of shots is limited.

\section{Theory} \label{sec:Theory}
A natural choice \cite{Gottesman1998Fault,Gottesman2001,Jena2019} for a qudit-based QC is to decompose the operator-to-be-measured $\hat{O}$ into $p$ Pauli Strings (PS)
\begin{subequations}
    \begin{align}
    \hat{O} = \sum_{i=1}^{p} c_i \hat{P}_i \quad \text{ with } \quad \hat{P}_i = \bigotimes_{j=1}^{q} \hat{X}_{d_j}^{r_j} \hat{Z}_{d_j}^{s_j}, \label{eq:PauliDecomposition}\\
    \hat{X}_{d_j}^{r_j} \hat{Z}_{d_j}^{s_j} = \sum_{\mu=0}^{d_j-1} \ket{(\mu + r_j)\%d_j} \omega^{s_j \, \mu} \bra{\mu}, \label{eq:DefGenPauli}
\end{align} \label{eq:PDecomposition}
\end{subequations}
where $c_i \in \mathbb{C}$ and the PS $\hat{P}_i$ are tensor products of $q$ Pauli operators (PO) $\hat{X}_{d_j}^{r_j} \hat{Z}_{d_j}^{s_j}$ acting on the $j$-th qudit with dimension $d_j$ and exponents $r_j,s_j \in \{0,\dots,d_j-1\}$.
The PO that make up the PS are products of rank $d_j$ operators $\hat{X}_{d_j}\ket{\mu} = \ket{(\mu+1) \% d_j}$ and $\hat{Z}_{d_j}\ket{\mu} = \omega_{d_j}^\mu \ket{\mu}$, where $\omega_{d_j} = \exp(2\pi\mathrm{i}/d_j)$ and $x \% d_j$ is the remainder of $x$ under division by $d_j$.
For $d = 2$, these are the standard Pauli $\hat{X}$ and Pauli $\hat{Z}$. 
Note that the PS constructed from these PO have eigenvalues $\omega_{d_P}^{\mu}$, $\mu = 0,\dots,d_P-1$, where $d_P = \text{lcm}(\vec{d)}$ is the dimension of the PS and $\vec{d} = \{d_1,d_2,\dots,d_q\}$. Therefore, $P_i^\dagger \neq P_i$ for $d_P > 2$. 
When $\hat{O}$ is hermitian, each $\hat{P}_i$ in the decomposition in Eq.~\eqref{eq:PauliDecomposition} will be accompanied by $\hat{P}_i^{\dagger}$.

Most (non-bosonic) observables in physical systems are expressed as tensor products of spin operators $\hat{S}_x$, $\hat{S}_y$, and $\hat{S}_z$ rather than the PO in Eq.~\eqref{eq:DefGenPauli}. To write $\hat{O}$ as in Eq.~\eqref{eq:PauliDecomposition}, it is therefore essential to understand how spin operators are expressed in terms of qudit PO. Following the standard definition (see App.~\ref{Sec:App_Spin}), a $d_S$-dimensional spin operator
\begin{align*}
    \hat{S}_{\alpha} = \sum_{r,s} c_{\alpha}^{r,s} \hat{X}_{d_S}^{r} \hat{Z}_{d_S}^{s} \quad \text{with} \quad \alpha = x,y,z
\end{align*}
is obtained from the coefficients
\begin{subequations}
    \begin{align}
        c_x^{r,s} 
        & = 
        \frac{1}{d_S}\sum_{\mu=0}^{d_S-2} \sigma_{\mu} \left( \frac{\delta(r-1)}{\omega^{s \mu}} + \frac{\delta(r+1-d)}{\omega^{s (\mu+1)}}\right)
        ,\label{eq:SpinDecompositionX}
        \\
        c_y^{r,s} 
        & = 
        \frac{\mathrm{i}}{d_S}\sum_{\mu=0}^{d_S-2} \sigma_{\mu} \left( \frac{\delta(r-1)}{\omega^{s \mu}} - \frac{\delta(r+1-d)}{\omega^{s(\mu+1)}}\right)
        ,\label{eq:SpinDecompositionY}
        \\
        c_z^{r,s} 
        & = 
        \frac{2}{d_S}\sum_{\mu=0}^{d_S-1} \left(\frac{d_S-1}{2} -\mu \right) \,  \omega^{-s \mu} \, \delta(r)
        ,\label{eq:SpinDecompositionZ}
    \end{align} \label{eq:SpinDecomposition}
\end{subequations}
where $\sigma_\mu = \sqrt{2((d_S+1)/2)(\mu+1)-(\mu+1)(\mu+2)}$ with spin $(d_S-1)/2$. Here, $\delta$ is the Kronecker delta with $\delta(0) = 1$ and $\delta(x) = 0$ for $x \neq 0$.
Notice that changing the basis from Spin to Pauli never results in an exponential increase of terms in the decomposition in Eq.~\eqref{eq:PDecomposition}. Indeed, for a given dimension $d_S$, the numbers of coefficients $c_x^{r,s}$, $c_y^{r,s}$, and $c_z^{r,s}$ in Eqs.~\eqref{eq:SpinDecomposition} are $2d_S$, $2d_S$, and $d_S$, respectively.
As such, each local operator $\hat{O}$ that has a polynomial number of Spin terms -- all physical observables -- will also have a polynomial number of PS.

We visualise the estimator $\widetilde{O} = \sum_{i} c_i \widetilde{P}_i \approx \langle \hat{O}\rangle = \sum_{i} c_i \hat{P}_i$ of $\hat{O}$ by constructing a graph representation of the observable, e.g. in Fig.~\ref{Fig:EstimationProcess}(a), where vertices (gray dots) correspond to PS and edges connect $\hat{P}_i$ to $\hat{P}_j$ if they commute $[\hat{P}_i,\hat{P}_j] = 0$. 
We assign weights $c_i \widetilde{P}_i$ and $c_i c_j \,\widetilde{\mathcal{C}}(\hat{P}_i,\hat{P}_j)$ [Eq.~\eqref{eq:ScaledCovarianceMatrix}] to vertices and edges (including self-edges), respectively [see Fig.~\ref{Fig:EstimationProcess}(a)]. The sum of the vertex weights is $\widetilde{O} \approx \langle \hat{O} \rangle$, while the edge weights contribute to the estimation error. We provide a more detailed explanation of this process in Sec.~\ref{sec:Algorithm}.

The central task of any measurement algorithm is to distribute $M$ shots (i.e., repeated measurements of the same quantum state) between the PS in Eq.~\eqref{eq:PDecomposition} such that the estimation variance $(\widetilde{\Delta O})^2$ (or estimation error $\sqrt{(\widetilde{\Delta O})^2}$) is minimized. Here,
\begin{subequations}
    \begin{align}
    (\widetilde{\Delta O})^2 &= \sum_{i,j=1}^{n} c_i c_j \,\widetilde{\mathcal{C}}(\hat{P}_i,\hat{P}_j), \label{eq:EstimationError} \\
    \widetilde{\mathcal{C}}(\hat{P}_i,\hat{P}_j) &= \frac{m_{ij} + 2}{(m_i + 2)(m_j + 2)} \widetilde{Q}^{(1,1)}_{ij} \label{eq:ScaledCovarianceMatrix},
\end{align} \label{eq:costfunction}
\end{subequations}
and $Q^{(A,B)}_{ij} = \langle (\hat{P}^A_i)^{\dagger} \hat{P}^B_j\rangle - \langle (\hat{P}^A_i)^{\dagger} \rangle \langle \hat{P}^B_j \rangle$ with $A,B \in \{0,1,\dots,d_P - 1\}$ being the elements of the covariance matrix. We use ``$\sim$'' to indicate estimators, such that $\widetilde{O}, \widetilde{P}_i$ and $\widetilde{Q}_{ij}^{(1,1)}$ are estimates, constructed from the measurement outcomes, for $\langle \hat{O} \rangle, \langle \hat{P}_i \rangle$ and $Q_{ij}^{(1,1)}$, respectively. Lastly, $m_{ij}$ denotes the number of shots where $\hat{P}_i$ and $\hat{P}_j$ were measured simultaneously, while $m_i$ is the number of times $\hat{P}_i$ was measured overall. 
Note that the estimated covariances $\widetilde{Q}_{ij}^{(1,1)}$ are non-zero if and only if the respective PS have been measured together, which can only be done when they commute. Distributing the shots to these groups of commuting PS [which we call ``cliques'', see green, red and blue connected vertices in Fig.~\ref{Fig:EstimationProcess}(a)] \cite{Shlosberg2023} is a complex task that our algorithm addresses, see Sec.~\ref{sec:Algorithm}.

Calculating $(\widetilde{\Delta O})^2$ in Eq.~\eqref{eq:EstimationError} such that it correctly reproduces the statistical variance of $\widetilde{O}$ requires a model that respects the physical constraints of the measurement process. 
When overlapping cliques [with shared PS, see vertex $j$ in Fig.~\ref{Fig:EstimationProcess}(a)] are measured, there is no well-established procedure to calculate $\widetilde{Q}^{(1,1)}_{ij}$. 
As discussed in \cite{Shlosberg2023}, this could lead to unphysical values of $(\widetilde{\Delta O})^2$ (e.g., negative). 
To avoid this problem, in \cite{Shlosberg2023} we intentionally overestimated the error contributions from the covariances $\widetilde{Q}_{ij}^{(1,1)}$. 
Here, we build a Bayesian model that can tackle qudit observables while simultaneously exploiting the physical constraints that commuting PS must satisfy. 
This yields more accurate error estimates that result in better allocation and consequently lower errors overall (see Sec.~\ref{sec:NumResults}). 
In the following section, we explain our Bayesian model.

\subsection{Bayesian Estimation Model} \label{subsec:BayesianModel}
\begin{figure*}
  \includegraphics[width=\textwidth]{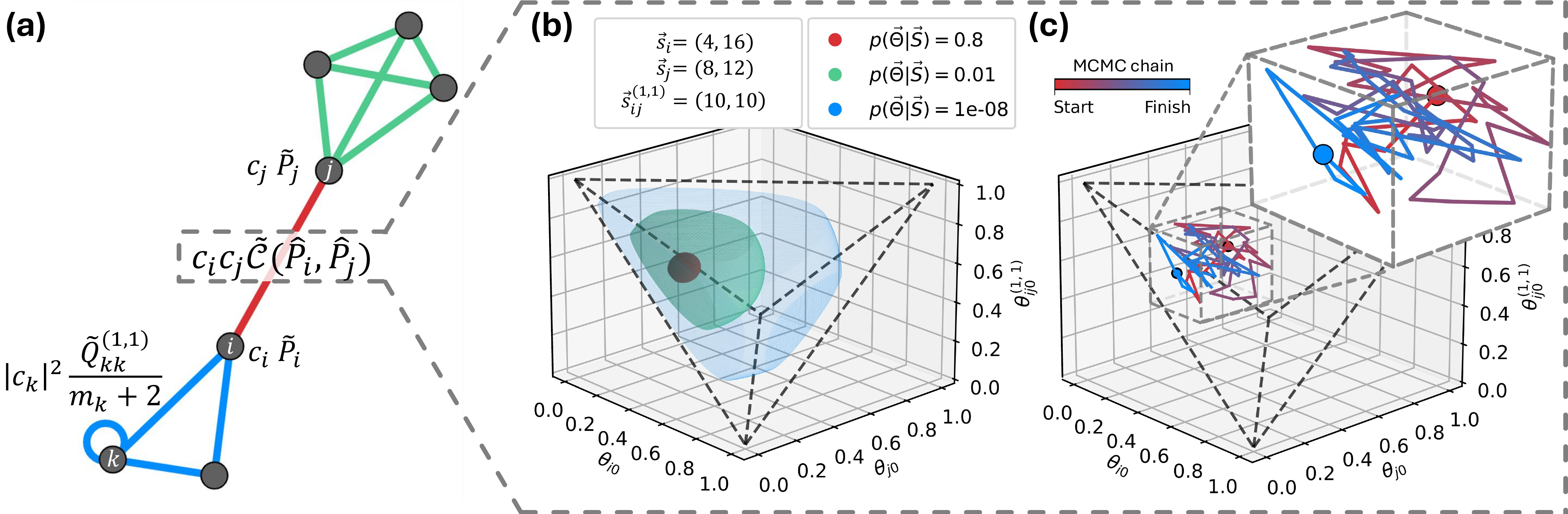}
  \caption{Bayesian estimation process for $d_P = 2$. \textbf{(a)} displays the commutation graph of an observable $\hat{O}$ with $p=7$ PS in its decomposition Eq.~\eqref{eq:PauliDecomposition}. Vertices represent PS, and are weighted with their estimated averages. Edges connect commuting PS, and are weighted by their error contribution $c_ic_j\,\widetilde{\mathcal{C}}(\hat{P}_i,\hat{P}_j)$ in Eqs.~\eqref{eq:costfunction}. Self-edges (for clarity omitted except for the $k$-th vertex) are labelled with the estimated variances $|c_k|^2 \widetilde{\mathcal{C}}(\hat{P}_k,\hat{P}_k) = |c_k|^2 Q^{(1,1)}_{kk}/(m_k+2)$. Therefore, $\widetilde{O}$ [$(\widetilde{\Delta O})^2$] is the sum over all vertices [edges, including self-edges]. As explained in Sec.~\ref{sec:Theory}, the estimations rely on a MCMC procedure. For the covariance contribution of PS $i$ and $j$,  \textbf{(b)} depicts the posterior in Eq.~\eqref{eq:posterior} (normalized such that $\text{max}(p) = 1$) by means of equipotential surfaces (light blue, green and red, see legend). The tetrahedron marked by the dashed lines is the boundary $R$ from Eq.~\eqref{eq:BoundConditions}. \textbf{(c)} shows one chain at the basis of the MCMC process. Its colour is mapped to represent the evolution of the chain during the MCMC procedure. In (b,c) we employed the measurement outcomes $\vec{s}_i$, $\vec{s}_j$ and $\vec{s}^{(1,1)}_{ij}$ reported in the legend.}
  \label{Fig:EstimationProcess}
\end{figure*}
Averages and variances of random variables with a finite number of discrete outcomes (such as a die or a PS) can be described by Bayesian statistics \cite{Gelman1995,Kruschke2014}.
The estimator of $f(\vec{\theta}_i)$, where $f$ is a function of the underlying probabilities $\vec{\theta}_i = \{ \theta_{i0},\theta_{i1},\dots,\theta_{i \,d-1} \}$ of the PS, is
\begin{subequations}
    \begin{align}
        \widetilde{f} = \int f(\vec{\theta}_i) \, p_i(\vec{\theta}_i|\vec{s}_i) \, \text{d}\vec{\theta}_i, \label{eq:singleEstimator}\\
        p_i(\vec{\theta}_i|\vec{s}_i) \propto \prod_{\mu=0}^{d_P - 1} \theta_{i\mu}^{s_{i\mu} + a_{i\mu}}. \label{eq:singlePosterior}
    \end{align} \label{eq:DefEstimator}
\end{subequations}
Specifically, $\theta_{i\mu}$ and $s_{i\mu}$ are the probability and number of times the outcome $\mu$ is obtained when measuring $\hat{P}_i$, respectively, and $\vec{a} = \{a_{i0},\dots,a_{i(d-1)}\}$ are priors \cite{Gelman1995,Kruschke2014}. In this work $a_{i\mu} = 1$ for all $i$ and $\mu$.
Employing Eqs.~\eqref{eq:DefEstimator}, we find $\widetilde{P}_i$ and $\widetilde{Q}_{ii}^{(1,1)}$ to be
\begin{subequations}
    \begin{align}
    &\widetilde{P}_i = \sum_{\mu=0}^{d_P - 1} \widetilde{\theta}_{i\mu} \omega^{\mu}_{d_P} \, \text{ with } \,
    \widetilde{\theta}_{i\mu} = \frac{s_{i\mu} + a_{i\mu}}{\sum_\mu a_{i\mu} + \sum_{\mu} s_{i\mu}}, \label{eq:EstimatorTheta} \\ 
    &\widetilde{Q}_{ii}^{(1,1)} = \sum_{\mu,\nu=0}^{d_P -1} \omega_{d_P}^{\mu} (\omega_{d_P}^{\mu} - \omega_{d_P}^{\nu}) \widetilde{\theta_{i\mu} \theta_{i\nu}} \, \text{ with} \label{eq:EstimatorVariance} \\
    &\widetilde{\theta_{i\mu}\theta_{i\nu}} = \frac{ (s_{i\mu} + a_{i\mu})(s_{i\nu} + a_{i\nu})}{(\Sigma_\mu a_{i\mu} + \Sigma_{\mu} s_{i\mu})(\Sigma_\mu a_{i\mu} + \Sigma_{\mu} s_{i\mu} +1)}.
    \end{align} \label{eq:EstimatorStandard}
\end{subequations}
Eq.~\eqref{eq:EstimatorTheta} suffices to estimate $\langle \hat{O}\rangle \approx \widetilde{O} = \sum_i c_i\widetilde{P}_i$, while Eq.~\eqref{eq:EstimatorVariance} includes the diagonal terms within the covariance matrix $\widetilde{Q}^{(1,1)}$ in Eq.~\eqref{eq:ScaledCovarianceMatrix}. Below, we explain how we determine $\widetilde{Q}_{ij}^{(1,1)}$ for $i\neq j$.

The following method for the improved estimation of the covariances $\widetilde{Q}_{ij}^{(A,B)}$ is one of the main contributions of this work. It both overcomes the limitations of the approach in AEQuO \cite{Shlosberg2023} and is at the basis of the enhanced accuracy  reported in our numerical investigations in Sec.~\ref{sec:NumResults}.
The key idea is recognizing that the product of two different PS, $(\hat{P}^A_i)^{\dagger}$ and $\hat{P}^B_j$, is again a PS $\hat{P}^{(A,B)}_{ij} = (\hat{P}^A_i)^{\dagger}\hat{P}^B_j$ with underlying probabilities $\vec{\theta}^{(A,B)}_{ij}$. Since $\hat{P}^{(A,B)}_{ij}$ can be measured every time $\hat{P}_i$ and $\hat{P}_j$ are simultaneously probed, these PS can be estimated without any additional resources.
When considering $\hat{P}^{(A,B)}_{ij}$ as its own Pauli, the covariance $Q_{ij}^{(A,B)}$ becomes a function of single PS averages and their probabilities $\vec{\theta}_{i}$, $\vec{\theta}_{j}$, and $\vec{\theta}^{(A,B)}_{ij}$.

While $\vec{\theta}_i$, $\vec{\theta}_j$ and $\vec{\theta}^{(A,B)}_{ij}$ are independent variables, they must respect physical constraints that depend on each others' values.
As an example, consider a qubit observable and assume $\theta_{i0} = \theta_{j1} = 1$. In this case, the probability $\vec{\theta}^{(1,1)}_{ij0}$ must be equal to $0$. However, for $\theta_{i0} = \theta_{j1} = 1/2$, $\vec{\theta}^{(1,1)}_{ij0}$ can vary between $0$ and $1$.
These constraints arise from the observation that the probability of simultaneously measuring the outcomes $\mu$ for $\hat{P}_i$ and $\nu$ for $\hat{P}_j$ must be positive and upper limited by one. 
By writing these probabilities in terms of the individual probabilities $\theta_{i\mu}$ and $\theta_{j\nu}$ as well as their covariances $Q_{ij}^{(A,B)}$ (see App.~\ref{sec:App_Boundary_Cond}), we find the $2d^2_P$ inequalities
\begin{align}
0 \, \leq \, \theta_{i\mu}\theta_{j\nu} + \frac{1}{d_P^2} \sum_{A,B=0}^{d_P-1} \omega^{A \mu - B \nu} Q^{(A,B)}_{ij} \, \leq \, 1. \label{eq:BoundConditions}
\end{align}
To obtain boundary conditions that depend on $\vec{\theta}_{i}$, $\vec{\theta}_{j}$ and $\vec{\theta}^{(A,B)}_{ij}$ we write $Q^{(A,B)}_{i,j} = \langle \hat{P}^{(A,B)}_{ij} \rangle - \langle (\hat{P}^A_i)^{\dagger} \rangle \langle \hat{P}^B_j \rangle = \sum_{\mu}\theta^{(A,B)}_{ij\mu} \omega_{d_P}^{\mu} - \sum_{\mu,\nu}\theta_{i\mu}\theta_{j\nu} \, \omega_{d_P}^{B\nu-A\mu}$.

Our approach for calculating the covariances between two PS prescribes
\begin{subequations}
    \begin{align}
    &
    \widetilde{Q}^{(1,1)}_{ij} = \int \limits_{R} Q^{(1,1)}_{ij} p(\bm{\vec{\Theta}}|\bm{\vec{S}}) \, \text{d}\bm{\vec{\Theta}},\label{eq:EstimationIntegral} \\
    &p(\bm{\vec{\Theta}}| \bm{\vec{S}}) \propto \prod_{\mu=0}^{d_P -1} \theta_{i\mu}^{s_{i\mu}} \, \theta_{j\mu}^{s_{j\mu}}  \,(\theta_{ij\mu}^{(1,1)})^{s_{ij\mu}^{(1,1)} }, \label{eq:posterior}
\end{align}\label{eq:EstimatorCovariance}
\end{subequations}
where $Q^{(1,1)}_{ij}$ is defined below Eq.~\eqref{eq:BoundConditions}, $R$ is the region bounded by Eq.~\eqref{eq:BoundConditions}, $\bm{\vec{\Theta}} = \{\vec{\theta}_i,\vec{\theta}_j,\vec{\theta}^{(1,1)}_{ij}\}$ and $\bm{\vec{S}} = \{\vec{s}_i,\vec{s}_j,\vec{s}^{(1,1)}_{ij}\}$. The posterior [Eq.~\eqref{eq:posterior}] is the product of the individual PS posteriors [Eq.~\eqref{eq:singlePosterior}], with priors $a_{i\mu}=a_{j\mu} = a_{ij\mu} = 1$ for all $\mu = 0,\dots, d_P - 1$.
The function $p(\bm{\vec{\Theta}}| \bm{\vec{S}})$ in Eq.~\eqref{eq:posterior} is depicted in Fig.~\ref{Fig:EstimationProcess}(b) for the case of qubits $d_P = 2$. The equipotential surfaces (red, green, blue) lie within the boundary $R$ delimited by dashed lines.
These result from the boundary conditions  $|1-\theta_{i0} - \theta_{j0}| \leq \theta^{(A,B)}_{ij0} \leq 1 - |\theta_{i0} - \theta_{j0}|$ [Eq.~\eqref{eq:BoundConditions}], and yield the tetrahedron in Figs.~\ref{Fig:EstimationProcess}(b-c).

Since the integrals in Eq.~\eqref{eq:EstimationIntegral} cannot be analytically calculated, we employ a Markov chain Monte Carlo (MCMC) approach \cite{Hastings1970,Newman1999} and develop a sampling procedure that satisfies the boundary conditions in Eq.~\eqref{eq:BoundConditions} by construction. 
The first step is to map the probabilities $\bm{\vec{\Theta}}$ to a quantum state represented by a point on the two-qudit generalization of the Bloch sphere. The second step is to enable the random walk at the basis of MCMC. The novelty is that we perform this random walk on the Hilbert space defined at step one, rather than the probabilities' space $\bm{\vec{\Theta}}$ on which $p(\bm{\vec{\Theta}}| \bm{\vec{S}})$ in Eq.~\eqref{eq:posterior} is defined. To do so, we mix the initial quantum state with a new state that is obtained from the first via a random Haar unitary. The mixing is regulated by our protocol (App.~\ref{Sec:App_MCMC}) such that the step of the random walk is of appropriate size to ensure convergence (see below). The third step is to determine, from each quantum state within the random walk, the corresponding probabilities $\vec{\theta}_{ij}^{(1,1)}$, $\vec{\theta}_{i}$ and $\vec{\theta}_{j}$ that are employed to estimate $\widetilde{Q}_{ij}^{(1,1)}$ in Eq.~\eqref{eq:EstimationIntegral}. This sampling process produces the MCMC chain shown in Fig.~\ref{Fig:EstimationProcess}(c). 

For the integration we utilize several independent MCMC chains that are initialized to the maximum of Eq.~\eqref{eq:posterior}, found using the mode of the Dirichlet distribution \cite{Johnson2000}. Once the Geweke \cite{Geweke1992} and the Gelman-Rubin diagnostics \cite{Gelman1992} verify convergence for the chains (or a maximum number of samples is reached), we stop and use the sampled probabilities to determine $\widetilde{Q}_{ij}^{(1,1)}$.
As can be seen in Fig.~\ref{Fig:EstimationProcess}(c) the chains sample closely to the region of highest probability in Fig.~\ref{Fig:EstimationProcess}(b).
A more detailed description of this process is available in App.~\ref{Sec:App_MCMC}.

\section{Algorithm} \label{sec:Algorithm}
We base the algorithm for shot allocation on AEQuO \cite{Shlosberg2023}, which is (to our knowledge) state-of-the-art for qubit observables. 
Here, we focus on the elements that enable the measurement of qudit systems and on the novelties introduced after \cite{Shlosberg2023}. 
We begin by decomposing the observable $\hat{O}$ into PS, see Eq.~\eqref{eq:PauliDecomposition}. 
The coefficients $c_i = \text{tr}(\hat{P}_i^{\dagger} \hat{O})/\prod_{i=1}^{q}d_i$ for $\hat{O}$ expressed as a sum of Spin operators are in Eqs.~\eqref{eq:SpinDecomposition}. 
From this decomposition, we construct a commutation graph [Fig.~\ref{Fig:EstimationProcess}(a)], where vertices (grey dots) correspond to PS and edges connect $\hat{P}_i$ to $\hat{P}_j$ if $[\hat{P}_i,\hat{P}_j] = 0$. 
At this point, we can choose whether to build the graph such that general commutation relations are allowed, or if we restrict to bitwise (see Secs.~\ref{Sec:Overview} and \ref{sec:NumResults}). We then determine a clique cover, where cliques group PS that can be simultaneously measured in one shot. 
In Fig.~\ref{Fig:EstimationProcess}(a) there are three cliques, with the red one having an overlap with both the blue (vertex $j$) and the green (vertex $i$).

To allocate the shots among these cliques, we assign weights $c_i \widetilde{P}_i$ and $c_i c_j \,\widetilde{\mathcal{C}}(\hat{P}_i,\hat{P}_j)$ to vertices and edges (including self-edges), respectively [Fig.~\ref{Fig:EstimationProcess}(a)]. 
The sum of the vertex weights is the estimate $\widetilde{O}$ of the expectation value $\langle \hat{O} \rangle$, while the edge weights build up the estimation variance $(\widetilde{\Delta O})^2$ in Eqs.~\eqref{eq:costfunction}. 
The contribution of individual cliques to $(\widetilde{\Delta O})^2$ can thus be evaluated by summing over all edges within that clique. 
We then allocate measurement shots iteratively to the cliques that provide the largest decrease in the estimation variance $(\widetilde{\Delta O})^2$ in Eqs.~\eqref{eq:costfunction}. 

The measurement of each clique relies on finding a Clifford circuit \cite{Gottesman1998Fault,Aaronson2004} that diagonalizes not only all PS within the clique but also their products $\hat{P}_{ij}^{(1,1)} = \hat{P}^{\dagger}_i \hat{P}_j$ (see Sec.~\ref{sec:Theory}). 
Following \cite{Jena2019}, our code includes a classically efficient subroutine that determines these circuits. 
When general [bitwise] commutation relations are used to build the graph in Fig.~\ref{Fig:EstimationProcess}(a), the classical algorithm that finds these circuits requires a time that is quadratic [linear] with respect to the qudit number $q$, and yields circuits whose depths are quadratic [equal to one] with respect to $q$. 
The Clifford circuits ensure that the eigenstates are elements of the computational basis, enabling the measurement of the non-hermitian PS $\hat{P}_i$. For the experimental results presented in Sec.~\ref{sec:ExpResults}, we use the generalized qudit Hadamard, Phase and generalized controlled-NOT gates \cite{Gottesman1998Fault} implemented with the pulse sequences as described in \cite{Ringbauer2022}

The outcomes in these diagonalized bases provide the data for the $\vec{s}_i$, $\vec{s}_j$ and $\vec{s}_{ij}^{(1,1)}$ employed by the Bayesian estimation in Sec.~\ref{sec:Theory}. 
After a user-specified number of measurement shots, the covariance graph is updated, with variances and covariances calculated as per Eqs.~\eqref{eq:EstimatorStandard} and \eqref{eq:EstimatorCovariance}. 
Updating the covariance matrix $\widetilde{Q}^{(1,1)}$ serves to better estimate each clique's error contribution and results in lower errors overall (see Sec.~\ref{sec:NumResults}).
This iterative shot-allocation process continues until the measurement budget is spent.

\section{Numerical Results} \label{sec:NumResults}
\begin{figure*}[!tbp]
  \includegraphics[width=1\textwidth]{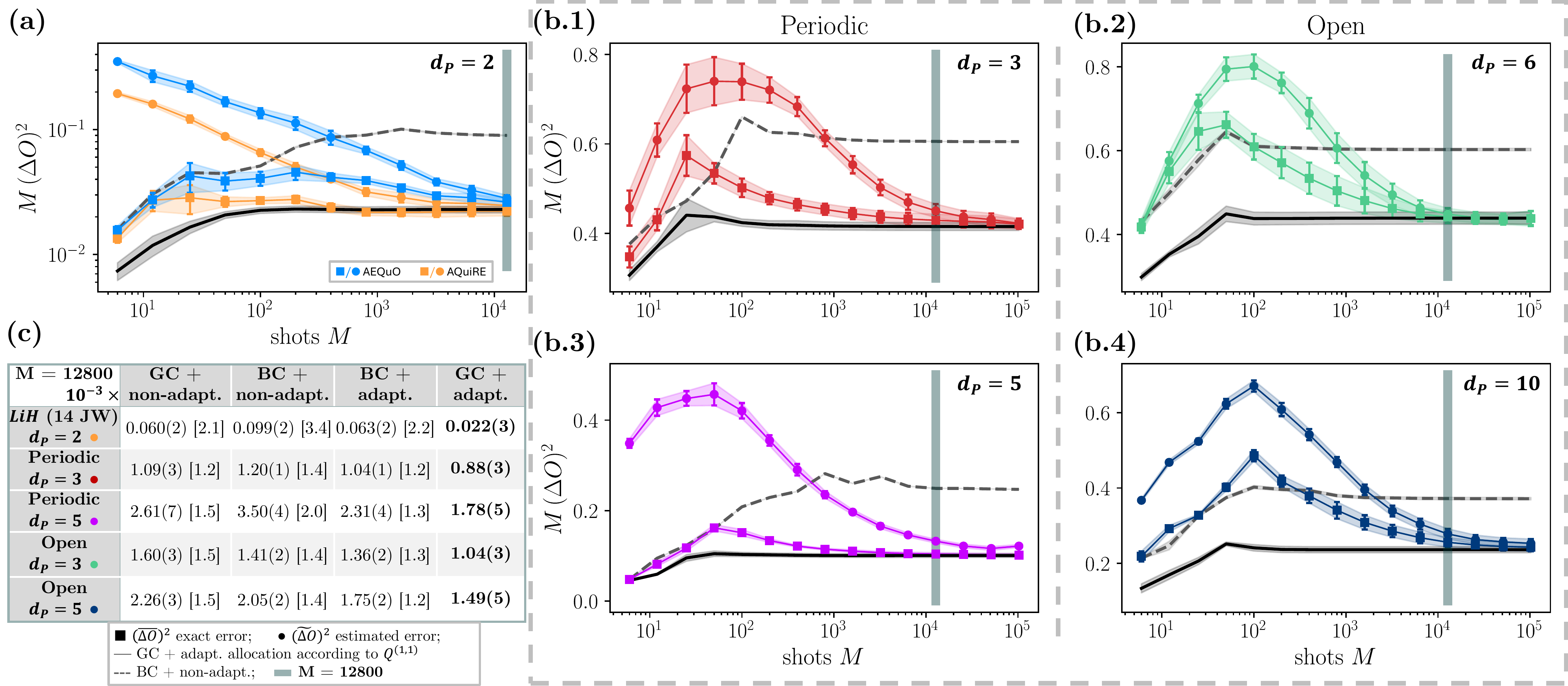}
  \caption{(a-b) Rescaled relative estimation variance $M(\Delta O)^2$ as a function of $M$ for ground state energies of various Hamiltonians. In the y-axes, $(\Delta O)^2$ is either $(\overline{\Delta O})^{2}$ (squares) or $(\widetilde{\Delta O})^{2}$ (circles) (see the legend and main text), divided by the exact, average $\langle \hat{O}\rangle ^2$. Points are the averages over $20$ classically simulated experiments and the error bars their standard deviations.
  Full coloured lines are obtained with \textbf{GC+adaptive} and without prior information on the input quantum state. AEQuO's \cite{Shlosberg2023} results are in light blue; all other colours are obtained with {\algo}. The black (dashed gray) line is the variance $(\overline{\Delta O})^2$ obtained for \textbf{GC+adaptive} (\textbf{BC+non-adaptive}) allocation according to the true covariance matrix $Q^{(1,1)}$(only the PS weights $c_i$).
  In (a), we use the \text{LiH} (JW) molecule \cite{Cao2019} to benchmark AEQuO against {\algo}.
  (b) demonstrates the performance of {\algo} for different LGT Hamiltonians \cite{Haase2021,Paulson2021,Meth2023}. The left (right) column considers periodic (open) boundary conditions and the top (bottom) row approximates the gauge fields with qutrits (ququints). Parameters for the open plaquette Hamiltonian, encoded on four qubits and one qudit ($d=3,5$) \cite{Haase2021,Paulson2021,Meth2023}, are $m=-50$, $\Omega = 5$ and $g^2 = 1$ \cite[Eq.~(1)]{Paulson2021}; the periodic plaquette Hamiltonian also uses $g^2 = 1$ \cite[Eq.~(4)]{Paulson2021} and is encoded on three qudits ($d=3,5$).
  (c) Averaged errors $\sqrt{(\widetilde{\Delta O})^{2}}$ (standard deviations in brackets, bold indicates lowest) obtained by {\algo} for Hamiltonians in (a) and (b).
  The simulations were performed with $8$ MCMC chains with upwards of $500$ samples and a tuned acceptance rate of $25\%$. For further details, see App.~\ref{Sec:App_MCMC}.} 
  \label{Fig:NumericalResults}
\end{figure*}

In the following, we compare the impact of two different settings (each with two options) that modify the allocation strategy.
Firstly, \textbf{GC} (general commutation)/\textbf{BC} (bitwise commutation) describes the commutation rule used to build the cliques. 
As \textbf{GC} is less restrictive than \textbf{BC}, the former generally yields more and larger cliques. 
Secondly, \textbf{adapt.}/\textbf{non-adapt.} describes if the covariance matrix is updated (or not) during the allocation process. 
For the \textbf{non-adapt.} setting, only the PS weights $c_i$ play a role in the allocation, while for the \textbf{adapt.} setting, $\widetilde{Q}^{(1,1)}$ is updated during the process, according to Sec~\ref{sec:Theory}.

Our numerical results are presented in Fig.~\ref{Fig:NumericalResults}. 
We demonstrate the performance of {\algo} using the $\text{LiH}$ molecule \cite{Cao2019} and 2D lattice gauge theory QED Hamiltonians with different qudit dimensions and boundary conditions \cite{Haase2021,Paulson2021,Meth2023}. 
In panels (a) and (b), we depict the progression of the exact $\mapsto$ squares [estimated $\mapsto$ circles] relative estimation variance $(\overline{\Delta O})^2$ [$(\widetilde{\Delta O})^2$]. 
Both $(\widetilde{\Delta O})^2$ and $(\overline{\Delta O})^2$ are calculated using Eqs.~\eqref{eq:costfunction}, where the latter substitutes $\widetilde{Q}_{ij}^{(1,1)} \mapsto \langle \hat{P}_i \hat{P}_j \rangle - \langle \hat{P}_i\rangle \langle \hat{P}_j\rangle$, i.e. uses the exact covariances that are experimentally unknown. 
While $(\widetilde{\Delta O})^2$ is directly computed from the (simulated) experimental outcomes, $(\overline{\Delta O})^2$ is the exact variance of $\widetilde{O}$, which we use here as a reference.
To contextualize our results, we employ two benchmarks. 
First, the variance $(\overline{\Delta O})^2$ obtained from allocating \textbf{adapt.} with \textbf{GC} according to the real covariance matrix $Q^{(1,1)}$ (solid black line). 
Second, the variance $(\overline{\Delta O})^2$ obtained when allocating \textbf{non-adapt.} with only \textbf{BC} (dashed gray line).
In the limit of many measurements $M \mapsto \infty$, these are lower and upper bounds, respectively, for the estimation variance $(\widetilde{\Delta O})^2$ obtained by our qudit measurement protocol. 

Although designed for qudit, it is also crucial to show that {\algo} outperforms state-of-the-art qubit protocols.
In Fig.~\ref{Fig:NumericalResults}(a), we use the qubit Hamiltonian of the LiH molecule to demonstrate the advantages of {\algo} over AEQuO \cite{Shlosberg2023}. As can be seen from the plot, both AEQuO and {\algo} converge to the black line. This means that with a sufficiently large measurement budget $M$ both approaches allocate measurements as if the covariances $Q^{(1,1)}_{ij}$ of the unknown input quantum states were known \textit{before} the measurement.
Yet, {\algo} yields variances $(\overline{\Delta O})^2$ and $(\widetilde{\Delta O})^2$ that are considerably smaller than the ones of AEQuO, especially for limited shot numbers $M$. 
This follows from the enhanced Bayesian estimation procedure in Sec.~\ref{sec:Theory}. 
Particularly at low $M$, {\algo} entails a better error estimation which in turn results in a better allocation and faster convergence to the optimal relative error (black line). 

Between $M=10$ and $M=300$ {\algo} (in orange) entails the same accuracy as AEQuO (blue) with roughly $40\%$ [$60\%$] less shots if we employ the estimated [exact] variances $(\widetilde{\Delta O})^2$ [$(\overline{\Delta O})^2$].
Similarly, for $M = 10^3$ [$M = 10^4$], the estimated variances (dots) are within $10\%$ and $15\%$ [$5\%$ and $10\%$] of the exact error (squares) for {\algo} and AEQuO, respectively.
This means that at $M \approx 10^3$ {\algo} yields exact errors that are comparable with the ones obtained with exact covariances (black line). In comparison, AEQuO requires $M \approx 10^4$. 

In Fig.~\ref{Fig:NumericalResults}(b) we test {\algo} with qutrits (b.1), ququints (b.3), mixed qubit-qutrit (b.2) and qubit-ququint (b.4) systems. 
For the benchmark, we consider the U(1) LGT Hamiltonians from \cite{Meth2023}, that are an excellent example for utility provided by measuring observables on a qudit QC \cite{Meth2023}. 
In the left column, we assume no matter and periodic boundary conditions \cite{Haase2021,Paulson2021}. 
This is the smallest instance of a LGT in more than one spatial dimension and a proof of principle for the demonstration of the running of the coupling \cite{Paulson2021}. 
In the right column, we investigate a plaquette with elementary charges that interact with gauge fields. 
With open boundary conditions, this is the smallest instance of a LGT where the (infinite dimensional) electric and magnetic fields cannot be traced out and must be directly simulated on the QC \cite{Haase2021}.

As in (a), both the exact $(\overline{\Delta O})^2$ and estimated $(\widetilde{\Delta O})^2$ variances obtained in the simulated experiment converge to the variances attained by {\algo} when the exact covariances are provided beforehand (black lines). 
After an initial transient phase ($M \lesssim 100$) where several PS are yet to be measured, we witness a faster-than $1/M$ scaling followed by a horizontal plateau that indicates reaching the steady state $\propto 1/M$. 
This behaviour, qualitatively identical to the one in Fig.~\ref{Fig:NumericalResults}(a), is typical of this approach and showcases how {\algo} learns the relevant properties of the quantum state to enhance the allocation of the remaining shots. 
By comparing the gray dashed and black solid lines in Fig.~\ref{Fig:NumericalResults}(b), we can estimate the budget increase when \textbf{BC+non-adapt.} is used instead of \textbf{GC+adapt.} This is inferred from the ratio of the asymptotic values $M \mapsto \infty$ between the two lines. As can be seen from the four panels, \textbf{BC+non-adapt.} requires $37\%$, $36\%$, $96\%$, and $38\%$ more budget if compared to \textbf{GC+adapt.}
This percentage is generally higher when larger Hamiltonian are considered, as witnessed in our examples ($p = 86$, $51$, $324$ and $79$ in panels (b.1-4), respectively).

The performance between the different settings \textbf{BC} vs. \textbf{GC} and \textbf{adapt.} vs. \textbf{non-adapt.} (see Sec.~\ref{sec:NumResults}) is further investigated in Fig.~\ref{Fig:NumericalResults}(c). 
This table presents the estimated errors $\sqrt{(\widetilde{\Delta O})^2}$ obtained when the different settings are employed. For each problem instance (row) considered in the table, we indicate the lowest error in bold.
As expected, these are always achieved by the \textbf{GC + adapt.} settings, which entails both larger cliques and better allocation of the measurement shots. 
To quantify this advantage, numbers in square brackets are the ratios between the errors $\sqrt{(\widetilde{\Delta O})^2}$ of the associated column and the \textbf{GC + adapt.} setting. 
These numbers are indicative of the extra resources that must be allocated, with the corresponding settings, to attain the same accuracy as for the \textbf{GC + adapt.} case. 
For instance, to achieve error $1.78\times 10^{-3}$ (\textbf{GC + adapt.}) for the periodic $d_P = 5$ case with \textbf{BC + non-adapt.}, we would need $12800*2 = 25600$ measurements: twice the ones employed by \textbf{GC + adapt.}.

\section{Experimental Results}\label{sec:ExpResults}

So far, we discussed the scenario in which there is no hardware noise, such that the estimation variance $(\widetilde{\Delta O})^2$ in Eqs.~\eqref{eq:costfunction} could be obtained assuming only statistical noise affects the experimental outcomes. However, for NISQ and error-mitigated devices \cite{Bluvstein2024,Ott2024,Qiao2025,Kim2023,Meth2023,Delaney2024,Meng2024}, hardware noise \emph{must} be dealt with when measurement data are interpreted. For that reason, we introduce a scheme that quantifies how much it affects the specific state and operators being measured. This means that, rather than a bound for the worst-case scenario, our scheme precisely quantifies the detrimental effects of noise on the measurement itself. For example, when measuring $\hat{Z}$ with respect to $\ket{+}$ with an apparatus that is subject to a depolarizing channel, {\algo} recognizes that this error contribution does not affect the estimates for $\widetilde{Z}$ and $(\widetilde{\Delta Z})^2$.

Error awareness is enabled by quantifying the deviation $\widetilde{O}_{\rm e} - \widetilde{O}$ from the estimate $\widetilde{O}$, resulting from errors that would produce the estimate $\widetilde{O}_{\rm e}$. The core idea is to perform a user-specified number of stabilizer measurements \cite{Chan2024} for each Clifford circuit \cite{Gottesman1998Fault,Aaronson2004} used by {\algo} (to diagonalize the PS that are simultaneously measured at each step, see Sec.~\ref{sec:Algorithm}). With the details presented in App.~\ref{Sec:App_ErrorAwareness}, this allows to estimate the probabilities $\xi_i$ of obtaining an error when the PS $\hat{P}_i$ are probed. Once these $\widetilde{\xi}_{i} \approx \xi_i$ are calculated (using the same Bayesian approach described in Sec.~\ref{sec:Theory}), we can add the systematic variance $(\widetilde{O}_{\rm e} - \widetilde{O})^2$ to the statistical one $(\widetilde{\Delta O})^2$ from Eqs.~\eqref{eq:costfunction} to obtain the noise-aware variance $(\widetilde{\Delta O}_{\rm e})^2$
\begin{subequations}\label{eq:phys_errors}
\begin{align}
    (\widetilde{\Delta O}_{\rm e})^2 
    &= 
    (\widetilde{\Delta O})^2 + (\widetilde{O}_{\rm e} - \widetilde{O})^2
    ,\label{eq:phys_errors_var}
    \\
    \widetilde{O}_{\rm e} - \widetilde{O}
    &\simeq
    \sum_{i}c_i \widetilde{\xi}_{i}\sum_{\mu} 
    \left(
    \widetilde{\theta}_{i\mu}-\frac{1}{d_P}
    \right) 
    \omega^{\mu}_{d_P}
    .\label{eq:phys_errors_dev}
\end{align}
\end{subequations}
Here, we assumed that when an error occurs due to hardware noise, the measurement yields one of the eigenvalues $\omega_{d_P}^\mu$ ($\mu = 0,\dots,d_P-1$) at random, hence the factor $1/d_P$ in Eq.~\eqref{eq:phys_errors_dev}. 
In App.~\ref{Sec:App_ErrorAwareness} we explain how to loosen this constraint to generalize Eq.~\eqref{eq:phys_errors_dev} to obtain an error bound that is valid for all kinds of hardware noise, including highly correlated ones. However, as suggested by our experimental results below, even if the noise cannot be exactly described by a depolarizing channel \cite{Meth2022Probing}, the assumption of random error is excellent and $(\widetilde{\Delta O}_{\rm e})^2$ computed via Eqs.~\eqref{eq:phys_errors} properly quantifies hardware noise.

\begin{figure}
    \centering
    \includegraphics[width=8.6cm]{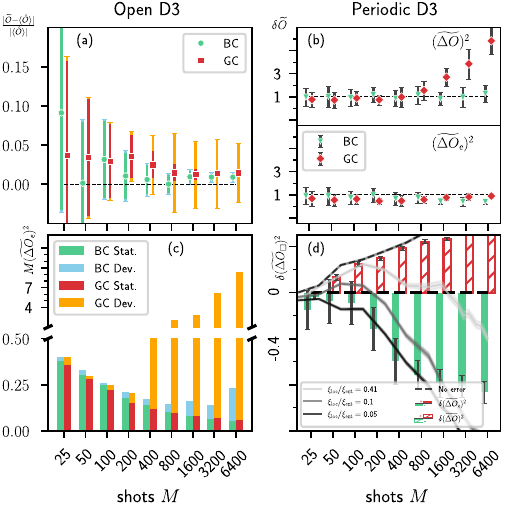}
    \caption{
    Experiment on the trapped-ion qudit QC \cite{Ringbauer2022}. The models for the left [right] column are the same as in Fig.~\ref{Fig:NumericalResults} except for the parameters $g^2 = 20$ [$g^2 = 0.01$] and input states $\ket{\psi} = \ket{1 0 1 0}_{\rm p} (\ket{0}_{\rm g} + \ket{1}_{\rm g})/\sqrt{2}$ [$\ket{\psi} = \otimes_{k=1}^{3} (\ket{0}_{\rm g}+\ket{1}_{\rm g}+\ket{2}_{\rm g})/\sqrt{3}$].
    \underline{Top row}: Relative single-shot (a) and average (b) distances $| \widetilde{O} - \langle \hat{O} \rangle | / | \langle \hat{O} \rangle |$ and $\delta \widetilde{O}$ (see main text), respectively, between experimental and exact results for both \textbf{BC+adapt.} (green) and \textbf{GC+adapt.} (red). Error bars in (a) are provided by {\algo} [colours as in panel (c)] while in (b) are standard deviations over $10$ runs.
    (c): Rescaled noise-aware variance $M(\widetilde{\Delta O}_{\rm e})^2$ splits into its statistical contribution $M(\widetilde{\Delta O})^2$ in green [red] and systematic contribution $M(\widetilde{O}_{\rm e} - \widetilde{O})^2$ in blue [orange] for the \textbf{BC+adapt.} [\textbf{G.C.+adapt.}] setting. 
    (d): Relative advantage $\delta (\widetilde{\Delta O}_{\square})^{2}$ -- see main text -- from using \textbf{GC} over \textbf{BC}. Filled and empty bars refer to $\delta (\widetilde{\Delta O}_{\rm e})^{2}$ and $\delta (\widetilde{\Delta O})^{2}$, respectively. Grey lines are numerically simulated runs with depolarizing noise affecting local and entangling gates with error rates $\xi_{\rm loc}$ and $\xi_{\rm ent}$, respectively (see App.~\ref{Sec:App_ErrorModels}). $\xi_{\rm loc} \approx 0.41 \% $ and $\xi_{\rm ent} \approx 7.9 \% $ are extracted from the experimental data; their ratio $\xi_{\rm loc} / \xi_{\rm ent} \approx 0.05$ is used for the darkest line. As in (b), error bars are standard deviations from $10$ repeated runs.
    }
    \label{Fig:ExpData}
\end{figure}

We show how {\algo}'s noise-awareness performs on the trapped-ion qudit QC thoroughly described in \cite{Ringbauer2022}. The basis states of the qudits are encoded in the electronic states of $^{40}\text{Ca}^+$ ions, which can be selectively addressed and manipulated by sequences of coherent laser pulses. In the following paragraphs, we compare two cases. First, the open plaquette Hamiltonian \cite{Haase2021,Paulson2021,Meth2023} in Fig.~\ref{Fig:NumericalResults}(b.2), encoded on four qubits (particles, suffix ``p'' below) and one qutrit (gauge field, ``g''). As input, we use $\ket{\psi_{\rm obc}} \propto  \ket{1010}_{\rm p} (\ket{0}_{\rm g} + \ket{1}_{\rm g})$, which has a fidelity to the ground state that exceeds $99\%$. Second, the periodic plaquette Hamiltonian from Fig.~\ref{Fig:NumericalResults}(b.1), encoded on three qutrits. Here, we use $\ket{\psi_{\rm pbc}} \propto \otimes_{k=1}^{3} (\ket{0}_{\rm g}+\ket{1}_{\rm g}+\ket{2}_{\rm g})$, i.e., the ground state in the magnetic regime of the model \cite{Haase2021}. Both $\ket{\psi_{\rm obc}}$ and $\ket{\psi_{\rm pbc}}$ have been chosen because they can be prepared with local operations only. Hence, the noise error {\algo} will calculate can be ascribed to the measurement process alone.

Our experimental results are presented in Fig.~\ref{Fig:ExpData}. 
The left and right columns include results pertaining to the open and periodic plaquettes, respectively. 
In the top row we show the absolute distance of the estimated average to the exact average of the Hamiltonian plotted against the number of shots $M$ for both \textbf{BC+adapt.} and \textbf{GC+adapt.}.
Panel (a) includes the result of a single measurement process; points (squares and circles) are $\lvert ( \widetilde{O} - \langle \hat{O}\rangle) / \langle \hat{O}\rangle \rvert$ and error bars $\sqrt{(\widetilde{\Delta O}_{\rm e})^2}/\lvert \langle \hat{O}\rangle \rvert$, see Eqs.~\eqref{eq:phys_errors}. In panel (b), on the other hand, we averaged $i=1,\dots,10$ measurement processes, each yielding $\widetilde{O}_i$, $(\widetilde{\Delta O})^{2}_{i}$ and $(\widetilde{\Delta O}_{\rm e})^{2}_{i}$, to obtain 
\begin{equation*}
    \delta\widetilde{O} 
    = 
    \frac{1}{10}
    \sum_{i=1}^{10}
    \frac{ 
    \left\lvert
    \widetilde{O}_i-\langle \hat{O} \rangle 
    \right\rvert
    }{
    \sqrt{(\widetilde{\Delta O}_{\square})^{2}_{i}}
    }
    .
\end{equation*}
In this last equation, ``$\square$'' distinguishes the upper plot from the lower plot, where we used $(\widetilde{\Delta O})^{2}_{i}$ and $(\widetilde{\Delta O}_{\rm e})^{2}_{i}$, respectively. The error bars in (b) are the associated standard deviations from the 10 repeated runs. 

The noise-awareness of {\algo} is further demonstrated in the bottom row of Fig.~\ref{Fig:ExpData}. 
In panel (c), we show how the rescaled variance $(\widetilde{\Delta O}_{\rm e})^{2}$ splits into a statistical part $M(\widetilde{\Delta O})^2$ (green and red for \textbf{BC} and \textbf{GC}, respectively) and a systematic part $M(\widetilde{O}_{\rm e} - \widetilde{O})^2$ (blue and orange), see Eqs.~\eqref{eq:phys_errors}. 
In (d), on the other hand, we display 
\begin{equation*}
    \delta (\widetilde{\Delta O}_{\square})^{2} 
    = 
    2\frac{
    (\widetilde{\Delta O}_{\square})^{2}|_{\rm BC} 
    - 
    (\widetilde{\Delta O}_{\square})^{2}|_{\rm GC}
    }{
    (\widetilde{\Delta O}_{\square})^{2}|_{\rm BC} 
    + 
    (\widetilde{\Delta O}_{\square})^{2}|_{\rm GC}
    },
\end{equation*}
i.e., a measurement of the relative advantage [$
\delta (\widetilde{\Delta O}_{\square})^{2} > 0
$] or disadvantage [$
\delta (\widetilde{\Delta O}_{\square})^{2} < 0
$] from \textbf{GC} over \textbf{BC}. This is calculated from the same data as Fig.~\ref{Fig:ExpData}(b) and, as before, ``$\square$'' distinguishes the case in which we used $(\widetilde{\Delta O})^{2}_{i}$ and $(\widetilde{\Delta O}_{\rm e})^{2}_{i}$. 

The upper row of Fig.~\ref{Fig:ExpData} shows that {\algo}, in a single measurement process [panel (a)], yields meaningful and reliable [panel (b)] estimates for both the average $\langle \hat{O} \rangle$ and its error. The error itself is \emph{not} an upper bound -- possibly orders of magnitude larger than $(\widetilde{\Delta O})^{2}$ -- but a direct estimation from the outcomes obtained within the measurement process. The role of hardware noise is also correctly captured by {\algo}. As can be seen in (a), for large values of $M$, $\widetilde{O}$ does not converge to $\langle \hat{O} \rangle$, but to another value that is shifted by erroneous measurement results. If we were employing $(\widetilde{\Delta O})^2$ (green or red error bars) instead of $(\widetilde{\Delta O}_{\rm e})^2$ (full bars, orange plus red or green plus blue), we would eventually produce a wrong estimate. This is particularly clear from the upper plot in panel (b), where for large $M$ the parameter $\delta \widetilde{O}$ exceeds unity -- meaning that $(\widetilde{\Delta O})^2$ fails in estimating the error. On the contrary, employing $(\widetilde{\Delta O}_{\rm e})^2$ as an error (lower plot) keeps $\delta \widetilde{O}$ around unity. This implies that {\algo} provides a statistically meaningful error, that is neither over- nor under-estimated. 

In Fig.~\ref{Fig:ExpData}(c) we show in detail how the statistical variance $(\widetilde{\Delta O})^2$ is the main contribution to the total noise-aware variance $(\widetilde{\Delta O}_{\rm e})^2$ for low shot numbers $M$, but becomes negligible at higher shot numbers, where the systematic variance $(\widetilde{O}_{\rm e} - \widetilde{O})^2$ dominates.
As in Fig.~\ref{Fig:NumericalResults}, we see an initial transient phase for the statistical variance $(\widetilde{\Delta O})^2$, where we witness a better than $1/M$ scaling. For larger $M$, $(\widetilde{\Delta O})^2$ attains a horizontal plateau that indicates reaching of the steady state $\propto 1/M$. On the contrary, aside from statistical fluctuation, the systematic variance $(\widetilde{O}_{\rm e} - \widetilde{O})^2$ is constant throughout the measurement process.

Panel (c) confirms that, when only considering the statistical variance $(\widetilde{\Delta O})^2$, \textbf{GC} can be advantageous -- see also Fig.~\ref{Fig:NumericalResults}.
We point out that, for $M \gtrsim 1600$, the statistical variance of \textbf{BC} is actually lower than that of \textbf{GC}: $(\widetilde{\Delta O})^2|_{\rm BC} < (\widetilde{\Delta O})^2|_{\rm GC}$.
This is an artifact from hardware noise, which is much higher for \textbf{GC} than for \textbf{BC} (see discussion below). An increase in noise results in a decrease of the estimated absolute values $|\langle \hat{P}_i \rangle|$ of each PS. Hence, the variances $(\Delta \hat{P}_{i})^2 = 1 - \langle \hat{P}_i \rangle^2$ become larger, leading to the increase in the statistical variance that can be seen in panel (c). Furthermore, we point out that between $50$ and $200$ shots, the \textbf{GC} scheme is advantageous despite the presence of hardware error. This indicates that even with large error rates of the entangling gates, the \textbf{GC} setting can be better in certain few-shot regimes; see also panel (d). 

Finally, in Fig.~\ref{Fig:ExpData}(d) we use the parameter $\delta (\widetilde{\Delta O}_{\square})^2$ (defined above) to compare the performances of the \textbf{BC} and \textbf{GC} strategies. There are two opposite phenomena, one favouring \textbf{GC} and the other \textbf{BC}. The first is that \textbf{GC} permits measuring more PS simultaneously compared to \textbf{BC}. The second is that \textbf{GC}, which relies on entangling gates for obtaining the measurement outcomes (see Sec.~\ref{sec:Algorithm} and \cite{Shlosberg2023,Jena2019}), is subject to higher hardware noise than \textbf{BC}. 

When we consider only the statistical error $\delta (\widetilde{\Delta O})^2$ (hashed bars), in accordance with the results in Fig.~\ref{Fig:NumericalResults}, the \textbf{GC} setting provides lower errors and is increasingly advantageous (except for $25$ shots, all hashed bars are red). 
However, with hardware noise $\delta (\widetilde{\Delta O}_{\rm e})^2$ (filled bars), this advantage is lost and the \textbf{BC} setting is on average better. Even so, we want to draw attention to the fact that the error bars in Fig.~\ref{Fig:ExpData}(d) indicate that there are runs where \textbf{GC} is advantageous in the regime of $M \lesssim 100$, something also observed in panel (c). On the other hand, for $M \gtrsim 100$, the hardware error of the entangling gates dominates and the \textbf{BC} setting is always the better choice. 
This suggests that, for sufficiently low entangling gate noise, a good strategy is to employ \textbf{GC} until the systematic error from hardware noise $(\widetilde{O}_{\rm e} - \widetilde{O})^2$ becomes comparable to the statistical noise $(\widetilde{\Delta O})^2$, and then switch to \textbf{BC}. This is readily implemented in {\algo}; it only requires changing the groups of PS that can be simultaneously measured during the measurement process (see Sec.~\ref{sec:Algorithm}).


An important point to address is then how reducing the hardware noise enhances the estimation accuracy. Thanks to {\algo}'s error awareness, we can directly answer this question. The gray lines in Fig.~\ref{Fig:ExpData}(d) are numerically simulated results in which we inject depolarizing noise \cite{Ramiro2023Enhancing} after each gate within the diagonalizing circuits employed for the measurements (see Sec.~\ref{sec:Algorithm} and App.~\ref{Sec:App_ErrorAwareness}). The darkest line is obtained by fitting the experimental data (bars), yielding an average local and entangling Clifford gate \footnote{
In this work, we employ the Hadamard, S, and sum gate as defined in \cite{Gottesman1998Fault}; see App.~\ref{Sec:App_CliffordGates}.
} 
error rates of $\xi_{\rm loc} = 0.41 \%$ and $ \xi_{\rm ent} = 7.9 \%$, respectively. Lighter lines are obtained by fixing $\xi_{\rm loc}$ and varying the ratio $\xi_{\rm loc} / \xi_{\rm ent}$, and the shadows correspond to standard deviations from $10$ repeated runs. The dashed line assumes $\xi_{\rm loc} = \xi_{\rm ent} = 0$, representing data as shown in Fig.~\ref{Fig:NumericalResults}.
The plot suggests that the regime of \textbf{GC} advantage scales roughly linearly with respect to $\xi_{\rm loc} / \xi_{\rm ent} \in [0.05,0.49]$. By decreasing $\xi_{\rm ent}$ by a factor of $2$ [$4$] (gray lines in the plot) the number of shots $M$ up to which $\delta (\widetilde{\Delta O}_{\rm e})^{2} \geq 0$ are $M \approx 200$ [$M \approx 1600$], demonstrating how much near-term algorithms can benefit from lowering the entangling gate's error rate $\xi_{\rm ent}$.

%

\section{Conclusion}\label{Sec:Conclusion}
We introduced {\algo}, an adaptive algorithm
to estimate both the mean and error of qudit quantum observables, as well as the influence of hardware noise in real-time.
At the heart of {\algo}, we developed a Bayesian model that can utilize the data collected from experiments with overlapping cliques of PS to reconstruct the estimators of the observable.
Through numerical simulations we show {\algo}'s advantage over other current state-of-the-art protocols. More interestingly, through implementation on a trapped-ion \emph{qudit} quantum computer we showcase the robustness of our algorithm with respect to hardware noise.
We demonstrate that employing entangling gates for the estimation, despite their increased noise rate, can be advantageous when the number of available shots is limited. 
We also quantify how an improvement in the quality of the entangling gates translates into an extended regime of enhanced accuracy enabled by the entangling gates. Specifically, we find roughly linear scaling of the measurement shots with respect to the ratio of entangling to local gate noise rate.

There are several ways one can extend the results presented in this work. On one hand, one can improve the strategy by which cliques are selected. By allowing the set of cliques to switch from general \textbf{GC} to bitwise \textbf{BC} commutation on the fly, one could utilize the advantage provided by entangling gates for low shot numbers and then use the obtained information to select bitwise commuting cliques (with lower noise), resulting in a faster convergence to the estimation accuracy limit set by hardware noise.
Alternatively, one can adapt {\algo} to minimize the total estimation variance $(\widetilde{\Delta O}_{\rm e})^2$ rather than the statistical variance $(\widetilde{\Delta O})^2$, thus allowing weighting the information provided by clique against the hardware noise of the circuit employed to measure it.
On the other hand, one can further improve the estimation directly, either through a Bayesian model that also utilizes information about non-commuting observables or, for the hardware noise awareness, by using pre-experiment available error tomography data as priors for the estimation.
Lastly, using the data collected by {\algo}, a logical next step would be not only to estimate the readout noise but also to correct for it.

\section*{Acknowledgments}
We thank Margaritis Voliotis for insightful discussion on Bayesian estimation. 
RS and LD acknowledge the EPSRC quantum career development grant EP/W028301/1 and the EPSRC Standard Research grant EP/Z534250/1. MM, PT and MR acknowledge support by the European Research Council (QUDITS, 101039522), and by the European Union under the Horizon Europe Programme--Grant Agreement 101080086--NeQST, by the Austrian Federal Ministry of Education, Science and Research via the Austrian Research Promotion Agency (FFG) hrough the projects FO999914030 (MUSIQ) and FO999921407 (HDcode) funded by the European Union-NextGenerationEU, and by the IQI GmbH. Views and opinions expressed are however those of the authors only and do not necessarily reflect those of the European Union or the European Research Council Executive Agency. Neither the European Union nor the granting authority can be held responsible for them. 
\bibliography{literature.bib}


\newpage
\appendix
\section{Spin-Operators in Higher Dimensions}\label{Sec:App_Spin}
In this work we use the following definitions of $d_S$ dimensional spin operators:
\begin{subequations}
    \begin{align}
    \hat{S}_x &= \sum_{\mu=0}^{d_S-2} (\ket{\mu+1}\bra{\mu} + \ket{\mu}\bra{\mu+1}) \sigma_{\mu}, \\
    \hat{S}_y &= \mathrm{i} \sum_{\mu=0}^{d_S-2} (\ket{\mu+1}\bra{\mu} - \ket{\mu}\bra{\mu+1}) \sigma_{\mu}, \\
    \hat{S}_z &= 2\sum_{\mu=0}^{d_S-1} \left(\frac{d_S-1}{2} - \mu\right)\ket{\mu}\bra{\mu},
\end{align} \label{eq:Definition_Spin_Matrices}
\end{subequations}
where $\sigma_{\mu} = \sqrt{2((d_S+1)/2)(\mu+1)-(\mu+1)(\mu+2)}$ with spin $(d_S-1)/2$. One finds the decomposition in Eqs.~\eqref{eq:SpinDecomposition} by substituting the spin operator definitions above into
\begin{align}
    c^{r,s}_\alpha = \frac{\text{tr}(\hat{P}_{r,s}^{\dagger} \hat{S}_\alpha)}{d_S},
\end{align}
where one can use the definition in Eq.~\eqref{eq:DefGenPauli} for the Paulis $\hat{P}_{r,s}$.

\section{Boundary Conditions} \label{sec:App_Boundary_Cond}
Here we explain how to derive the boundary conditions in Eq.~\eqref{eq:BoundConditions}. As described in Sec.~\ref{sec:Theory}, we use that the product of two PS forms a PS which we call $(\hat{P}^A_i)^{\dagger}\hat{P}^B_j \mapsto \hat{P}^{(A,B)}_{ij}$. Although the probabilities $\vec{\theta}_i$, $\vec{\theta}_j$, and $\vec{\theta}_{ij}$ are independent--meaning they cannot be uniquely determined from one another (see example in Sec.~\ref{sec:Theory}) -- their respective domains are affected by the specific values. Typically one would assume that the only conditions affecting $\vec{\theta}_{ij}$ are $0\leq \theta_{ij\mu} \leq 1$ and $\sum_\mu \theta_{ij\mu} = 1$. However, in order to obtain a physically accurate description of the system we need to account for the fact that the \textit{joint probabilities} $\vartheta_{i\mu j\nu}$ -- the probabilities of measuring the outcomes $\mu$ for $\hat{P}_i$ and simultaneously $\nu$ for $\hat{P}_i$ -- also need to be well defined $0\leq \vartheta_{i\mu j\nu} \leq 1$. 

To account for these $2d_{P}^2$ inequalities we first need to write them with respect to $\vec{\theta}_i$, $\vec{\theta}_j$ and $\vec{\theta}_{ij}$. We remind ourselves that the covariance with respect to these variables is
\begin{align}
\begin{split}
    Q^{A,B}_{i,j} &= \langle \hat{P}^{(A,B)}_{ij} \rangle - \langle (\hat{P}^A_i)^{\dagger} \rangle \langle \hat{P}^B_j \rangle \\ &= \sum_{\mu}\theta^{(A,B)}_{ij\mu} \omega_{d_P}^{\mu} - \sum_{\mu,\nu}\theta_{i\mu}\theta_{j\nu} \, \omega_{d_P}^{B\nu-A\mu}.
\end{split}
\end{align}
Since this representation of the covariance needs to align with the one using $\vartheta_{i\mu j\nu}$, for each choice of $A,B \in \{0,...,d_P-1\}$, we find $d_{P}^2$ equations
\begin{align}
    \sum_{\mu,\nu} \omega_{d_P}^{-A\mu}\omega_{d_P}^{B\mu} \vartheta_{i\mu j\nu} = Q^{A,B}_{i,j} + \sum_{\mu,\nu} \omega_{d_P}^{-A\mu} \theta_{i\mu} \omega_{d_P}^{B\nu} \theta_{j\nu}. 
\end{align}
By solving them for $\vartheta_{i\mu j\nu}$ we find
\begin{align}
    \vartheta_{i\mu j\nu} = \theta_{i\mu}\theta_{j\nu} + \frac{1}{d_P^2} \sum_{A,B=0}^{d_P-1} \omega^{A \, \mu - B \, \nu} Q^{(A,B)}_{ij},
\end{align}
leading directly to the inequalities shown in Eq.~\eqref{eq:BoundConditions}.

\section{Details on MCMC Integration} \label{Sec:App_MCMC}
We will now outline how we employ Markov chain Monte Carlo (MCMC) integration to calculate the integrals in Eq.~\eqref{eq:EstimationIntegral}. Specifically, we utilize importance sampling based on the Metropolis-Hastings (MH) algorithm \cite{Newman1999}, as this method does not scale with the number of integration variables. In general, MCMC integration relies on randomly sampling points ($\bm{\vec{\Theta}}$) in the integration volume (the region $R$ given by Eq.~\eqref{eq:BoundConditions}) and then averaging the values of the integrand [$Q^{(1,1)}_{ij} p(\bm{\vec{\Theta}}|\bm{\vec{S}})$] at the sampled points.
Importance sampling modifies this procedure by sampling not uniformly but according to a specifically chosen distribution [$D(\bm{\vec{\Theta}})$]. Here, the natural choice for this distribution is $D(\bm{\vec{\Theta}}) = p(\bm{\vec{\Theta}}|\bm{\vec{S}})$, simplifying the calculation to
\begin{align}
\begin{split}
    \widetilde{Q}^{(1,1)}_{ij} &= \int \limits_{R} Q^{(1,1)}_{ij} p(\bm{\vec{\Theta}}|\bm{\vec{S}}) \, \text{d}\bm{\vec{\Theta}} \\ &\approx \frac{1}{N}\sum_{k=0}^{N} \frac{Q^{(1,1)}_{ij} p(\bm{\vec{\Theta}}_k|\bm{\vec{S}})}{D(\bm{\vec{\Theta}}_k)} \\ &= \frac{1}{N}\sum_{k=0}^{N} Q^{(1,1)}_{ij}(\bm{\vec{\Theta}}_k)
\end{split}
\end{align}
where $N$ is the number of samples used in the integration.
This implies that we still only need to average the covariances $Q^{(1,1)}_{ij}$ for given $\bm{\vec{\Theta}}_k$, sampled according to $p(\bm{\vec{\Theta}}|\bm{\vec{S}})$ within $R$.

We now use the MH algorithm to generate these samples. We initialize the samples by finding the maximum of $p(\bm{\vec{\Theta}}|\bm{\vec{S}})$ within $R$ under the additional constraints that $(\vec{\theta}_i)_0$ and $(\vec{\theta}_j)_0$ are given by $\vec{\widetilde{\theta}}_{i}$ and $\vec{\widetilde{\theta}}_{j}$ from Eq.~\eqref{eq:EstimatorTheta}. To get all further samples $\bm{\vec{\Theta}}_k$ we need a symmetric proposal distribution $g(\bm{\vec{\Theta}}'|\bm{\vec{\Theta}}_k)$ that suggests a new sample $\bm{\vec{\Theta}}'$ that is either accepted or rejected with probability $\text{max}\{1,p(\bm{\vec{\Theta}}'|\bm{\vec{S}})/p(\bm{\vec{\Theta}}_k|\bm{\vec{S}})\}$. For that reason we developed a symmetric sampling strategy that gives normalized and usable samples $\bm{\vec{\Theta}}'$ even at high measurement numbers where $p$ is very sharp and automatically complies to the boundary conditions in Eq.~\eqref{eq:BoundConditions}.

\begin{figure}[!htbp]
    \centering
    \includegraphics[width=0.6\linewidth]{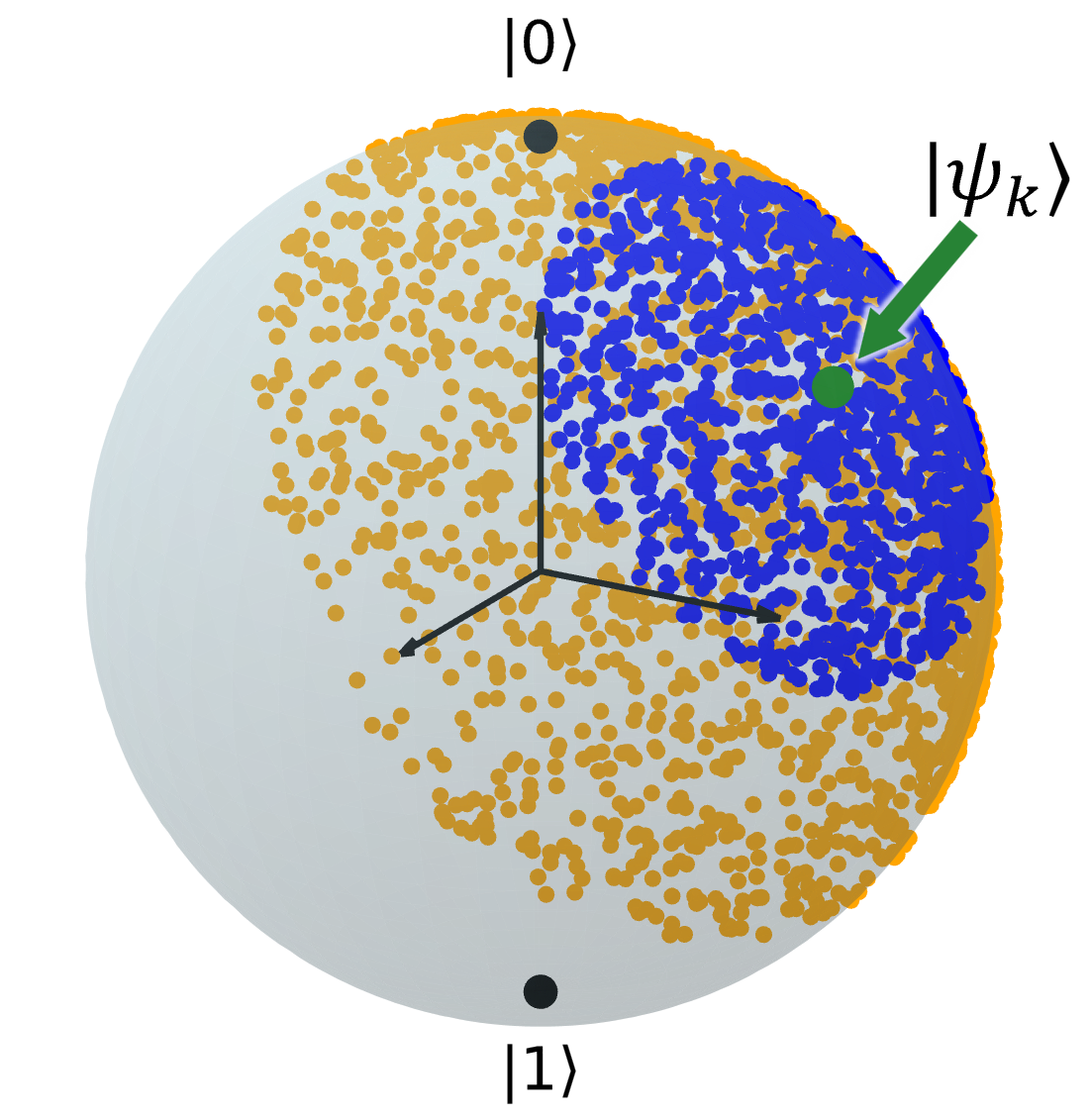}
    \caption{Visualization of the sampling process on a Bloch sphere. The green point is the initial state $\ket{\psi_k}$ and the blue (orange) points are possibilities for proposal states $\ket{\psi'}$ for $\gamma=0.95$ ($\gamma = 0.85$).}
    \label{Fig:Bloch_visualization}
\end{figure}

The main idea of our strategy consists of sampling not directly the probabilities but instead mapping these probabilities to the space of physical two-qudit states $\ket{\psi}$ and sampling there. This way we circumvent the issues of normalization and staying within the boundary conditions.
For a given starting sample $\bm{\vec{\Theta}}_k$ our method produces a proposal sample via the following steps. First, we map $\bm{\vec{\Theta}}_k$ to a corresponding two-qudit state $\bm{\vec{\Theta}}_k \mapsto \ket{\psi_k} = \sum_{i,j=0}^{d_P -1} \phi^{(k)}_{ij} \ket{i}\ket{j}$ (if we do not have the state already from the previous step) such that
\begin{align}
    \begin{split}
        (\theta_{i\mu})_k &= \sum_{j=0}^{d_P -1} |\phi_{\mu j}^{(k)}|^2 \\
        (\theta_{j\mu})_k &= \sum_{i=0}^{d_P -1} |\phi_{i \mu}^{(k)}|^2 \\
        (\theta^{(A,B)}_{ij\mu})_k &= \sum_{\,(B\cdot j' - A\cdot i')\% d_P = \mu} |\phi_{i' j'}^{(k)}|^2.
    \end{split} \label{eq:prob_mapping}
\end{align}
While this state can not be visualized using the Bloch-sphere Fig.~\ref{Fig:Bloch_visualization} gives an intuition for how this sampling is supposed to be understood with a state on the Bloch-sphere as an example. 

Next, we randomly sample another two-qudit state $\ket{\chi}$ according to the Haar distribution. We obtain the state of our new sample by constructing the following mixture between $\ket{\chi}$ and \ket{\psi_k}
\begin{align}
    \ket{\psi'} \propto \gamma \ket{\psi_k} + \sqrt{1-\gamma^2}\ket{\chi} \label{eq:mixture}
\end{align}
where $\gamma$ is the mixing parameter that determines the distance of the new state from the old one. In Fig.~\ref{Fig:Bloch_visualization}, $\gamma$ dictates the maximum distance of the blue/orange points $\ket{\psi'}$ from the green starting point \ket{\psi_k}. The closer $\gamma$ is to $1$, the smaller the distance is between the states. Before calculating the integral, we tune $\gamma$ such that the average acceptance rate of new samples is just above $25\%$. Phenomenologically we find that 
\begin{align}
    \gamma = 1 - \frac{1}{\text{min}\{\sum_\mu s_{i\mu}, \,\sum_\mu s_{j\mu}, \,\sum_\mu s^{(1,1)}_{ij\mu}\}}
\end{align}
usually satisfies this condition ($\gamma =0$ if no measurements have been taken).

Since we sample $\ket{\chi}$ uniformly, the probability of finding \ket{\psi_k} given $\ket{\psi'}$ is the same as finding $\ket{\psi'}$ given \ket{\psi_k}, meaning that our sampling is symmetric $ g(\bm{\vec{\Theta}}'|\bm{\vec{\Theta}}_k) = g(\bm{\vec{\Theta}}_k|\bm{\vec{\Theta}}') $.
Finally, we map the new state back to the probability space $\ket{\psi'} \mapsto \bm{\vec{\Theta}}'$ using Eq.~\eqref{eq:prob_mapping}. The resulting sample $\bm{\vec{\Theta}}'$ is now properly normalized (because $\ket{\psi'}$ is), fulfills the boundary conditions (because they need to be fulfilled for a physical state) and is accepted (on average) at least $25\%$ of the time.

\section{Error-awareness} \label{Sec:App_ErrorAwareness}
Here we will now go into detail on the derivation of Eq.~\eqref{eq:phys_errors} and how it can be used to calculate worst-case error bounds. We will also elaborate on how to perform the stabilizer measurements that build the basis of this error-awareness.

First, let us assume that the probability of obtaining a wrong outcome when measuring a certain PS $\hat{P}_i$ is given by $\xi_i$. What is now the expected deviation from the real $\langle \hat{O}\rangle$ due to such an error?
The average without errors $\langle \hat{O}\rangle_{\rm ne}$ is known
\begin{align}
    \hat{O} = \sum_{i} c_i \hat{P}_i \quad \Rightarrow \quad \langle \hat{O}\rangle_{\rm ne} = \sum_i c_i \sum_\mu \omega_{d_P}^{\mu} \theta_{i\mu}
\end{align}
but if we have errors with probability $\xi_i$, the average ``with error'' $\langle\hat{O}\rangle_{\rm e}$ shifts to
\begin{align}
    \langle\hat{O}\rangle_{\rm e} = \sum_{i} c_i \sum_{\mu} \omega_{d_P}^{\mu} [(1-\xi_i)\theta_{i\mu} + \xi_i \vartheta_{i\mu}], \label{eq:average_with_error}
\end{align}
where $\vartheta_{i\mu}$ is the probability of measuring the outcome $\mu$ when errors occur.

We can now rewrite Eq.~\ref{eq:average_with_error} to find the explicit difference between the two averages
\begin{equation}
    \langle\hat{O}\rangle_{\rm e} - \langle \hat{O}\rangle_{\rm ne} = \sum_{i} c_i \xi_i \sum_{\mu} \omega_{d_P}^{\mu}(\vartheta_{i\mu} - \theta_{i\mu}). \label{eq:true_deviation}
\end{equation}
For clarity, here we assume that the error is depolarizing in nature, which means that it results in each outcome with the same probability $\vartheta_{i\mu} = 1/d_{P}$. The estimator of Eq.~\ref{eq:true_deviation} becomes [see also Eq.~\eqref{eq:phys_errors}]
\begin{equation}
    \widetilde{O}_{\rm e} - \widetilde{O}_{\rm ne}\sim \sum_{i}c_i \widetilde{\xi}_{i}\sum_{\mu} (\widetilde{\theta}_{i\mu}-\tfrac{1}{d_P}) \omega^{\mu}_{d_P}
\end{equation}
where $\widetilde{O}_{\rm e} \approx \langle\hat{O}\rangle_{\rm e}$ and $\widetilde{O}_{\rm ne} \approx \langle\hat{O}\rangle_{\rm ne}$.

While the assumption of depolarizing noise $\vartheta_{i\mu} = 1/d_{P}$ is generally excellent, as it reflects the fact that errors on observables with $\theta_{i\mu} = 1/d_{P}$  do not affect the estimation, if information of the device specific $\vartheta_{i\mu}$ is known it can be used to significantly enhance the estimation accuracy.
Alternatively, for fixed $\widetilde{\theta}_{i\mu}$ and $\widetilde{\xi}_{i}$, one can vary $\vartheta_{i\mu}$ to find the true error bounds of the estimate
\begin{align}
    \max_{\vec{\vartheta}} \{\widetilde{O}_{\rm e} - \widetilde{O}_{\rm ne}\} \mapsto \text{error bound}
\end{align}
under constraints $\sum_{\mu} \vartheta_{i\mu} = 1, \, \forall i$.

To estimate $\widetilde{\xi}_i$ we use stabilizer measurements of modified versions of the circuits used in the experiment. By counting the frequency at which these measurements yield an error, we can develop a dataset $\vec{\zeta}_i = \{\zeta_{i,{\rm e}},\zeta_{i,{\rm ne}}\}$, wherein $\zeta_{i,{\rm e}}$ ($\zeta_{i,{\rm ne}}$) denotes the quantity of error-affected (error-free) measurements of $\hat{P}_i$. We may then employ the Bayesian principles described in Sec.~\ref{sec:Theory} to estimate the underlying unknown probability $\xi_i$.
To perform the stabilizer measurements we simplify the circuits $\hat{C}_\alpha$ by removing the single-qubit Hadamard gates (through adding three additional Hadamard, as $\hat{H}^4 = \hat{I}$). This way, the circuits take computational states to computational states, minimizing potential preparation error, while not reducing the error rate of the circuit. For the measurement, we select a random computational state $\ket{\psi^{\rm ideal}_{\rm out}}$ and find the computational state $\ket{\psi^{\rm ideal}_{\rm in}}$ that produces it, given the circuit $\hat{C}_\alpha$. We prepare $\ket{\psi^{\rm ideal}_{\rm in}}$ experimentally and apply $\hat{C}_\alpha$, recording an error if $\ket{\psi^{\rm exp}_{\rm out}}$ and $\ket{\psi^{\rm ideal}_{\rm out}}$ do not match.

\section{Clifford Gates} \label{Sec:App_CliffordGates}
The definitions for the Hadamard, Phase and controlled-NOT (CNOT) gate are readily known and universally agreed upon for the qubit case. For qudits, we follow the definitions in \cite{Gottesman1998Fault}. Then the Hadamard transform generalizes to the $d$-dimensional discrete Fourier transform 
\begin{align}
    \hat{H}_d\ket{j} \mapsto \sum_{i=0}^{d-1} \omega^{j\cdot i}\ket{i},
\end{align}
the $d$-dimensional phase gate is given by
\begin{align}
    \hat{S}_d\ket{j} \mapsto \omega^{j(j-1)/2}\ket{j},
\end{align}
and the CNOT gate yields the controlled-SUM (CSUM) gate 
\begin{align}
    \ket{i}\ket{j} \mapsto \ket{i}\ket{(i+j) \text{ mod } d}.
\end{align}
While the Hadamard and the Phase gate are realized by coherent laser pulses only coupling to a single ion at a time, the CSUM gate is implemented via the M\o lmer-S\o rensen (MS) interaction \cite{Sorensen_1999, Sorensen_2000}. Here, two qudits are simultaneously illuminated by a bichromatic light field, coupling the electronic states of the ions with their collective motion in the trap. For $d = 3$, the CSUM is realized by the sequence
\begin{equation*}
\begin{split}
    \text{CSUM} &= H_3^{(i)^{-1}}H_{2, (1, 2)}^{(j)} \\
    &\hspace{3em}H_{2, (1, 2)}^{(i)}\text{MS}^{(i,j)}_{(1,2)} H_{2, (1, 2)}^{(i)} \\
    &\hspace{9em}H_{2, (1, 2)}^{(j)}H_3^{(j)},\\\\  
\end{split}
\end{equation*}
where the superscripts label the addressed qutrit(s) $i$ and $j$, while the subscripts denote the dimensions and -- where applicable -- which basis states are coupled by the gate. 
Our apparatus enables single-qudit fidelities of $\mathcal{F}>99.9\%$, however, the average fidelity of the MS gate is given by $\mathcal{F}\approx 98.5\%$ \cite{Ringbauer2022}. Therefore, we identify the MS gate -- and consequently the CSUM gate -- as the major contribution to the device error estimation.

\section{Error models} \label{Sec:App_ErrorModels}
For the numerical simulations in Fig.~\ref{Fig:ExpData}(d) we employ a simple error model to simulate the effect of noisy gates on the allocation. First, we simulate the result as we did for Fig.~\ref{Fig:NumericalResults}, using the real state and diagonalization unitaries. Next, we calculate the error probability. Here, we assume that all local gates have the error probability $\xi_{\rm loc}$ and all two-qudit entangling gates the probability $\xi_{\rm ent}$, with an overall detection error rate of $\xi_{\rm detect}$. Assuming that the errors are independent from the state or other gates, the error probability $\xi(\hat{C}_i)$ of a given circuit $\hat{C}_i$ is the complement of the probability that no errors occur
\begin{equation}
\begin{split}
    \xi(C_i) = 1 - &\left[(1-\xi_{\rm detect}) (1- \xi_{\rm ent})^{\#_{\rm ent}(C_i)} \right. \\ 
      & \times\left.(1- \xi_{\rm loc})^{\#_{\rm loc}(C_i)} \right].
\end{split}
\end{equation}
Here, $\#_{\rm loc}$ and $\#_{\rm ent}$ denote the number or local and entangling gates, respectively.
Lastly, given the simulation result and the error probability of the corresponding diagonalization circuit, we let an error occur with probability $\xi(C_i)$, completely randomizing the simulation outcome. If no error occurs, the simulation result remains unchanged. 

To obtain these error rates from the measurement data, we again utilize a Bayesian model. If we denote the circuits used for the error-awareness stabilizer measurements with $C_i$ and the amount of erroneous and error-free measurements with $\mu_{i-}$ and $\mu_{i+}$, respectively, we can write the posterior, given a certain data set $(C_i,\mu_{i\pm})_{\forall i}$, as
\begin{align}
    p(\vec{\xi}|(C_i,\mu_{i\pm})_{\forall i}) &\propto \prod_{i} \xi(C_i)^{\mu_{i-}} [1-\xi(C_i)]^{\mu_{i+}}.
\end{align}
Thus we can estimate the means and variances of $\xi_{\rm detect}$, $\xi_{\rm ent}$ and $\xi_{\rm loc}$ as described in Sec.~\ref{sec:Theory}. 

\end{document}